\pdfoutput=1 
\documentclass[
superscriptaddress,
aps,
reprint,
prb,
floatfix,
]{revtex4-1}

 \makeatletter
\def\@fnsymbol#1{\ensuremath{\ifcase#1\or \dagger\or * \or \ddagger\or
   \mathsection\or \mathparagraph\or \|\or **\or \dagger\dagger
   \or \ddagger\ddagger \else\@ctrerr\fi}}
    \makeatother

\usepackage{braket}
\usepackage{textcomp}
\usepackage{gensymb}
\usepackage{physics}
\usepackage{multirow}
\usepackage{graphicx}
\usepackage{dcolumn}
\usepackage{bm}
\usepackage{amsmath}
\usepackage{color}
\usepackage{float}
\usepackage{array,url,kantlipsum}
\usepackage{verbatim}
\usepackage{xcolor}
\newcolumntype{P}[1]{>{\centering\arraybackslash}p{#1}}
\usepackage{tabularx}
\usepackage{array}
\newcolumntype{M}[1]{>{\centering\arraybackslash}m{#1}}

\def\JGU{\small{Johannes Gutenberg-Universit{\"a}t Mainz, Helmholtz-Institut Mainz, GSI Helmholtzzentrum f{\"u}r Schwerionenforschung, 55128 Mainz, Germany}}
\def\CAS{\small{Beijing National Laboratory for Condensed Matter Physics, Institute of Physics, Chinese Academy of Sciences, Beijing 100190, China}}
\def\UCB{\small{Department of Physics, University of California, Berkeley, California 94720, USA}}
\def\DLUT{\small{School of Control Science and Engineering, Dalian University of Technology, Dalian 116024, China}}

\begin{document}
\preprint{APS/123-QED}

\title{Temperature shift  of magnetic-field-dependent photoluminescence features of nitrogen-vacancy ensembles in diamond}

\author{Irena Rodzo\'n}
\thanks{These authors contributed equally to this work.}
\affiliation{\JGU }
\author{Xue Zhang}
\thanks{These authors contributed equally to this work.}
\affiliation{\JGU}
\affiliation{\DLUT}
\author{Viktor Iv{\'a}dy}
\affiliation{\it Department of Physics of Complex Systems, ELTE E{\"o}tv{\"o}s Lor{\'a}nd University, Egyetem t{\'e}r 1-3, H-1053 Budapest, Hungary}
\affiliation{MTA-ELTE Lend{\"u}let “Momentum” NewQubit Research Group, P{\'a}zm{\'a}ny P{\'e}ter, S{\'e}t{\'a}ny 1/A, 1117 Budapest, Hungary}
\affiliation{Department of Physics, Chemistry and Biology, Link{\"o}ping University, 581 83 Link{\"o}ping, Sweden}
\author{Huijie Zheng\footnotemark[1]}
\email{hjzheng@iphy.ac.cn}
\affiliation{\CAS}

\author{Arne Wickenbrock}
\affiliation{\JGU}
\author{Dmitry Budker}
\affiliation{\JGU}
\affiliation{\UCB}

 \date{\today}

\begin{abstract}

Recently significant attention has been paid to magnetic-field-dependent photoluminescence (PL)  features of the negatively charged nitrogen-vacancy (NV) centers in diamond. These features are used for microwave-free sensing and are indicative of the spin-bath properties in the diamond sample.
Examinating the temperature dependence of the PL features allows to identify both temperature dependent and independent features, and to utilize them in diamond-based quantum sensing and dynamic nuclear polarization applications.
Here, we study the thermal variability of many different features visible in a wide range of magnetic fields. To this end, we first discuss the origin of the features and tentatively assign the previously unidentified features to cross relaxation of NV center containing multi-spin systems. The experimental results are compared with theoretically predicted temperature shifts deduced from a combination of thermal expansion and electron-phonon interactions. A deeper insight into the thermal behavior of a wide array of the features may come with important consequences for various applications in high-precision NV thermometry, gyroscopes, solid-state clocks, and biomagnetic measurements. 

\end{abstract}

\maketitle

\section{Introduction}

Negatively charged nitrogen-vacancy (NV$^{-}$) centers in diamond are widely used for quantum sensing. These centers are sensitive to magnetic field, electric field, temperature, rotations, and strain, which enables the development of magnetometers, electric sensors, gyroscopes, thermometers, and multisensors\,\cite{DegenRMP2017,GeorgiosFrontiner2021, AshokPRA2012,Lourette_PRApplied_2023, Kucskonature2013, Neumannnanoletter2013}. Typically, NV-based sensors utilize optically detected magnetic resonance (ODMR) techniques involving microwave transitions between the ground-state Zeeman sublevels, see, for example, Ref.\,\cite{OortPRB1990}.
 Microwave-free techniques have also been developed recently \cite{WickenbrockAPS_2016,zheng2022electrical} based on cross-relaxation and energy level anti-crossing-related features in the magnetic field dependence of light-induced photoluminescence of the center \cite{jakobi2017, ZhengPRA_2020}.

For a sensor to reliably report a value of the measured parameter (e.g., magnetic field), it is important to understand and account for the effect of other parameters (e.g., temperature) to separate the effects of the two and, ideally, measure multiple parameters at the same time. With an eye on this, in this work, we investigate the temperature dependence of several cross-relaxation features used for microwave-free magnetometry and dynamic nuclear polarization applications. 

The zero-field splitting (ZFS) of the ground-state electronic $m_s=0$ and $m_s\pm1$ sublevels was found to have a nonlinear dependence on temperature \cite{acosta2010temperature, Chen_APL_2011,doherty2014temperature}, that was analyzed theoretically \cite{doherty2014temperature,Tang2023_JPCL}. The theoretical models identify the mechanism of the temperature dependence of the ZFS (along with those of the optical ZPL and strain-induced splittings) as a combination of thermal expansion and electron-phonon interactions.

\begin{figure} 
\includegraphics[width=\columnwidth]{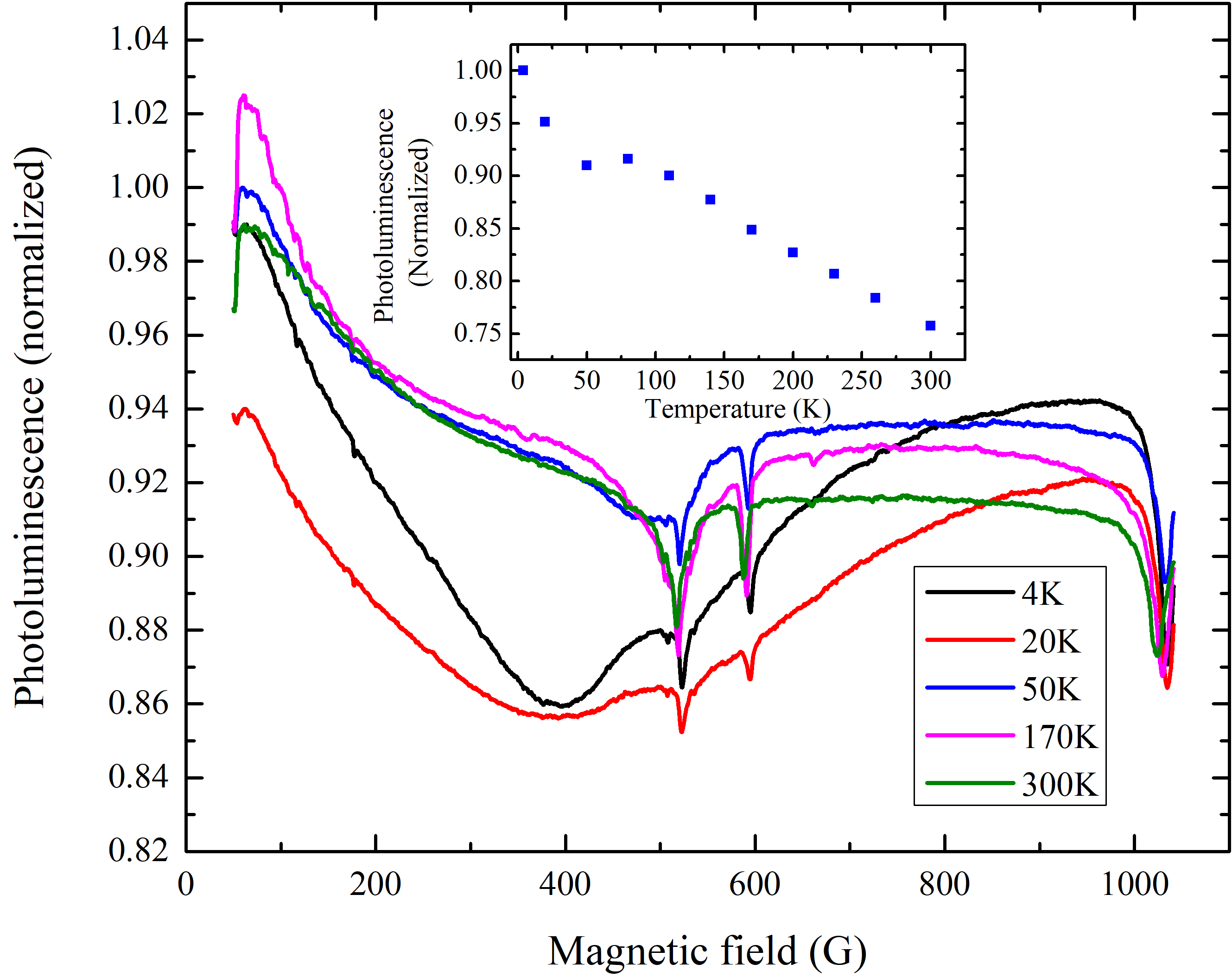} 
\caption {Temperature dependence of the magnetic field spectrum with E6 sample. The photoluminescence is normalized and vertically shifted for clarity. The inset shows the changes of photoluminescence with temperature at a magnetic field of around 800\,G. 
} \label{fig:plsignal} 
\end{figure}

\begin{figure}
\includegraphics[width=\columnwidth]{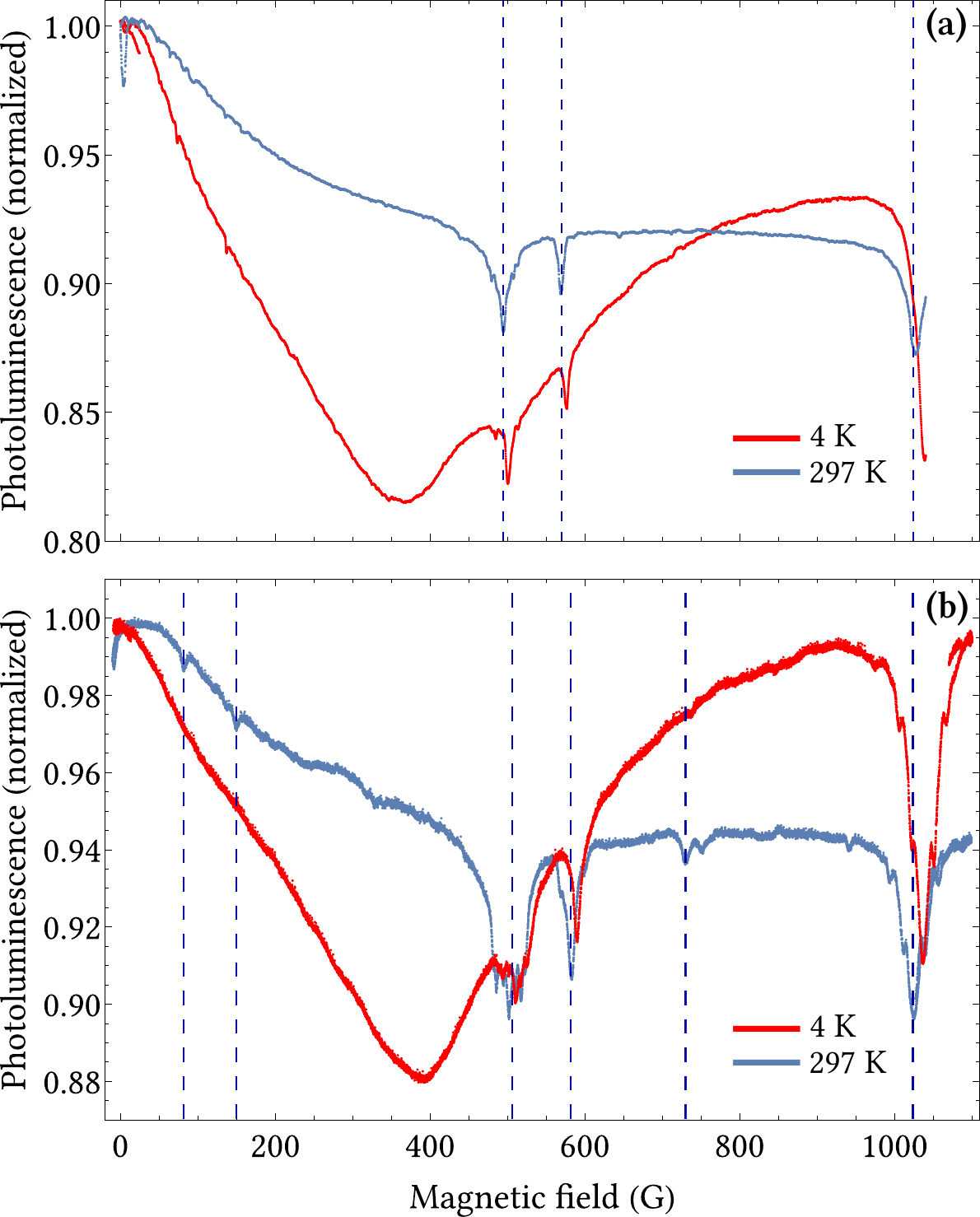}
\caption{
Photoluminescence for (a) E6 sample and (b) S2 diamond sample at room and cryogenic temperature. The dashed lines indicate investigated features, discussed in more detail in later sections. The significant shift of the feature near 1000\,G is due to the local slope of the curve.
}
\label{fig:plT_2samples}
\end{figure}

Recording the red/near-infrared photoluminescence intensity from a diamond sample with a high concentration of nitrogen-containing color centers (including NV$^-$) as a function of the magnetic field along one of the crystallographic axes, one observes a pattern (Figs.\,\ref{fig:plsignal} $\&$ \ref{fig:plT_2samples}) with a number of sharp features (see also Ref.\,\cite{Wunderlich2021}), the origin of most of which can be traced to cross-relaxation between NVs of different orientations or NVs and other paramagnetic centers such as P1 \cite{armstrong2010}. 

In addition to the sharp features, the photoluminescence also shows a broader (several hundred gauss) feature centered at zero  field. This is due to the NV$^-$ centers with axes that are not parallel to the magnetic field (i.e., ``off-axis NV centers''). The transverse component mixes electron-spin sublevels and reduces the efficiency with which the NV centers are optically pumped into the ``bright'' ground state with zero projection on the axis, thus reducing the photoluminescence \cite{armstrong2010, Rogers_2008}. Also of note is the behavior of the photoluminescence curves as a function of temperature.  The drop  of photoluminescence towards 400\,G at low temperatures is related to the dynamics in the excited state and was recently discussed in 
\cite{happacher2023PRL}, 
\footnote{The work of Ref.\,\cite{happacher2023PRL} examined single NV centers and there are notable differences between NV centers in different strain environments. Of note is that our samples show similar behavior although they were synthesized with different methods (E6--CVD,S2-HPHT), which generally produce different strain environment.}.

In this paper, we study the temperature dependence of the cross-relaxation features in the magnetic spectrum in the temperature range of 4--300\,K. We present the frequency shifts of each cross-relaxation feature and compare them with theoretical predictions. Features with relatively low-temperature dependence are observed, pointing to potential sensing applications.  In addition, we develop a scheme to efficiently and reliably predict cross-relaxation features of various multi-spin systems that we later employ to assess the origin of the magnetic field-dependent PL features. This in turn helps us to predict the temperature shift of the features and underpin our temperature-dependence measurements.

\section{Theory}

We begin by listing the observed sharp features in the magnetic-field dependence of the photoluminescence and discuss their origin, see Table\,\ref{Tab:predcition}. In order of increasing magnetic field, we focus on features appearing at the magnetic fields of 93\,G, 163\,G, 342\,G,
 512\,G (P1 center), 591\,G, 732\,G, 954\,G, and 1024\,G (the GSLAC), respectively. 
 As can be seen in Table \ref{Tab:predcition}, with the exception of the feature at 93\,G and 163\,G,
  all other features can be associated with various cross-relaxation conditions in the system, including cross-relaxation among the NVs of different orientations, NV-P1 cross-relaxation, as well as NV-$^{13}$C related processes. For the methodology used to predict the magnetic field values of possible PL features and for an in-depth analysis of the considered multi-spin systems, see Appendix B and Appendix C, respectively.

As the initial step in predicting the temperature dependence of the positions of the cross-relaxation features, we assume that the only parameter in the Hamiltonian that changes with temperature is the NV zero-field-splitting (ZFS) parameter $D$. This is motivated by the fact that the temperature dependence of the ODMR resonances is well described by this assumption \cite{doherty2014temperature}.

In the presence of a magnetic field at an angle with the NV axis, the eigenstates of the NV Hamiltonian are superpositions of magnetic sublevels with the amplitude of each sublevel depending on the magnitude and angle of the magnetic field with respect to the NV axis. We therefore diagonalize the corresponding NV Hamiltonian to find the eigenenergies as a function of $D$ and, correspondingly, of $T$. The results of this analysis are predictions of the temperature dependence, Table \ref{Tab:predcition}, that we compare with the experimental data below. We note that there are significant differences predicted for different features.

\begin{table*}[htb]
\caption{List of cross-relaxation features,  their origin, and the derivative of their temperature-dependent shift at 300\,K. A discussion of the considered multi-spin systems can be found in Appendix C. 
\label{Tab:predcition}}
\centering
\begin{tabular}{ M{2cm} | M{1cm}| M{3.5cm} | M{5cm} |M{2cm} | M{1.5cm} | M{1.5cm}}
  
\hline
Feature & B-field & \multicolumn{2}{|c|}{Origin} & Predicted slope at 300\,K  & Contrast 300\,K  & Contrast 4\,K \\ &
(G) & \multicolumn{2}{|c|}{} & (G/K) & (\%) & (\%) \\
\hline
GSLAC & 1024 & \small{
Avoided crossing between the $\ket{m_s=-1}$ and $\ket{m_s=0}$ states of the NV$^-$ center, in the presence of nuclear hyperfine couplings, strain, local fields, etc. 
}
 &
\begin{minipage}{.25\textwidth}
      \includegraphics[width=\linewidth]{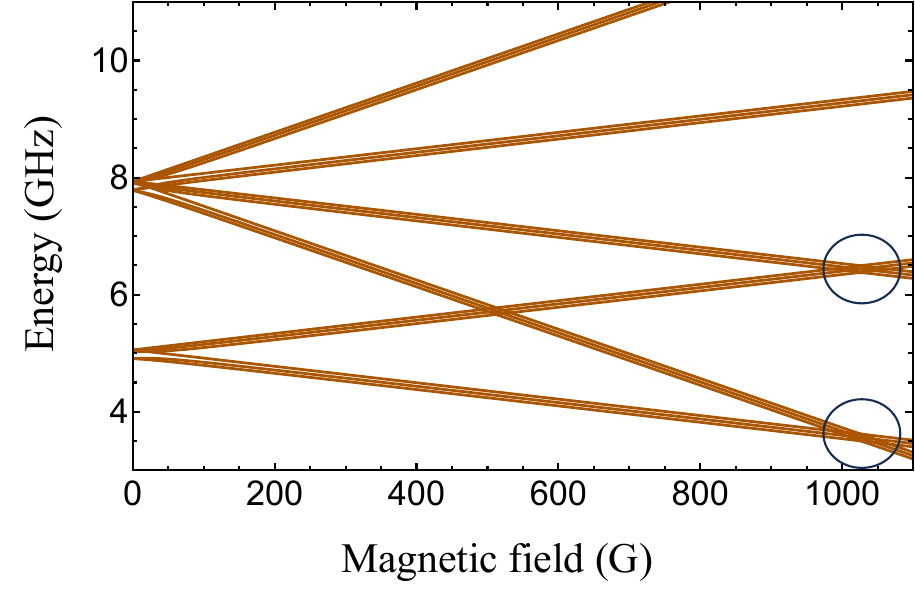}
    \end{minipage}
& -0.022 & 3.84 & 6.74 \\
\hline
NV$^-$-NV$^-$ - $^{13}$C  & 954 & \small{
Avoided crossing between the $\ket{m_s=-1}$ and $\ket{m_s=0}$ state of NV$^-$ centers interacting with a strongly coupled $^{13}$C nuclear spin} 
&
\begin{minipage}{.25\textwidth}
      \includegraphics[width=\linewidth]{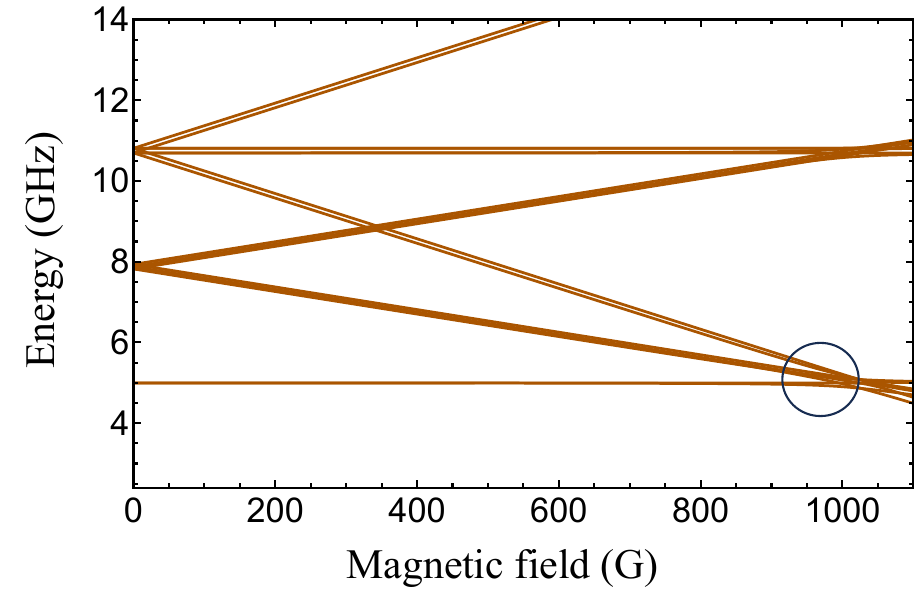}
    \end{minipage}
& -0.025 & - & -\\
\hline
2 $\times$ off-axis NV$^{-}$ - P1 & 732 & \small{Energy level crossing between the mixed states of two off-axis NV$^{-}$ centers interacting with a P1 center.} & \begin{minipage}{.25\textwidth}
      \includegraphics[width=\linewidth]{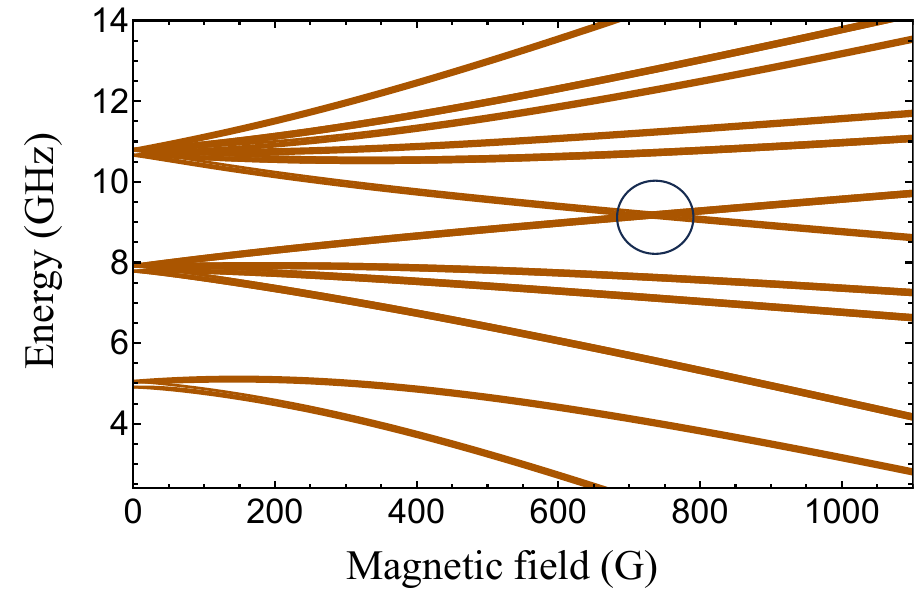}
    \end{minipage} & -0.017 & 0.67 & 0.16 \\
\hline
NV$^-$-NV$^-$ & 591 & 
Energy level crossing of NV$^-$ centers along different axes.
& 
\begin{minipage}{.25\textwidth}
      \includegraphics[width=\linewidth]{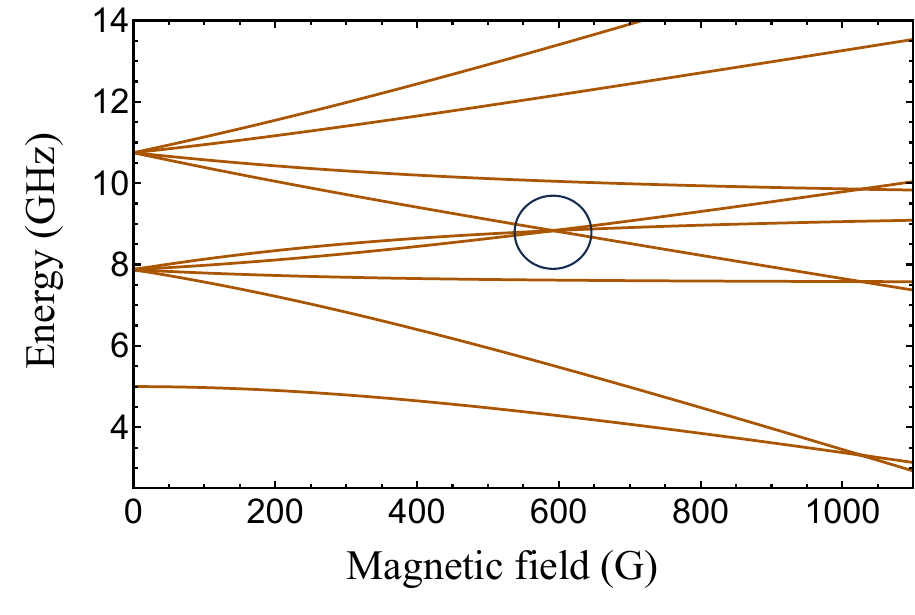}
    \end{minipage}
& -0.014 & 2.70 & 2.27 \\
\hline
NV$^-$-P1 & 512 & 
Avoided crossing between mixed NV$^{-}$, P1 center, and $^{14}$N nuclear spin states.
& 
\begin{minipage}{.25\textwidth}
      \includegraphics[width=\linewidth]{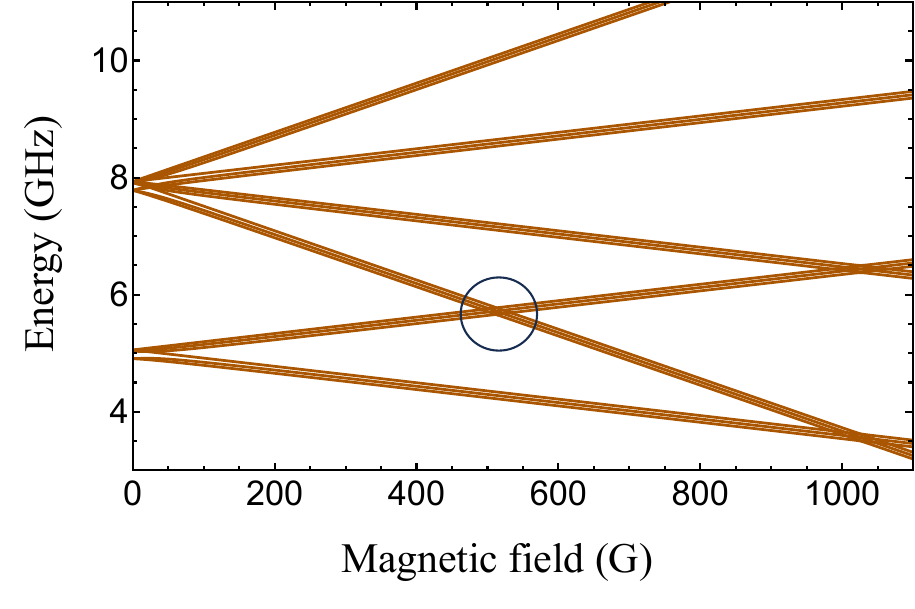}
    \end{minipage}
& -0.011 & 2.06 & 1.90 \\
\hline
 NV$^-$ - 2$\times$P1 & 342 & 
 Avoided crossing between the spin states of an NV$^-$ center, two P1 centers, and their nuclear spins.
  & 
\begin{minipage}{.25\textwidth}
      \includegraphics[width=\linewidth]{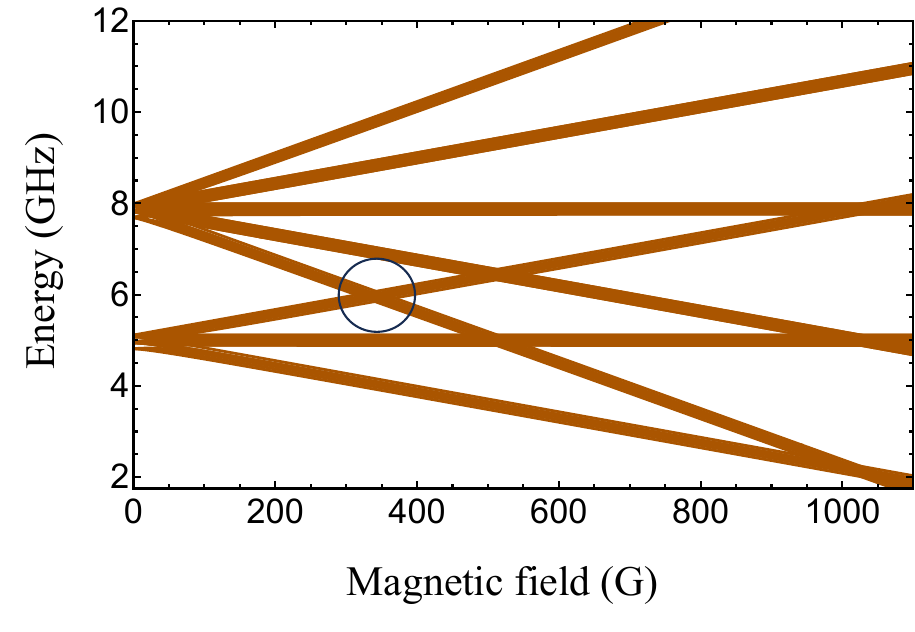}
    \end{minipage}  & -0.008 & - & -\\
\hline
-&163 & unknown 
& - & -& 0.24 & 0.07 \\
\hline
-& 93 & unknown 
& - & -& 0.26 & 0.06 \\
\hline
\end{tabular}
\end{table*}

A general Hamiltonian $H$ for the ground state can be written as:

\begin{equation}
\label{eq:Hamiltonian}
 H = H_{\rm{z}}+H_{\rm s}\,,
\end{equation}
where $H_{\rm{z}}$ represents Zeeman energy Hamiltonian and $H_{\rm s}=H_{\rm ee}+H_{\rm N}+H_{\rm {P1}}$ is spin-spin interaction that includes spin interaction of electron-electron $H_{\rm ee}$, of electron-nuclei $H_{\rm N}$, and of P1 centers $H_{\rm {P1}}$. The full expression of those Hamiltonians can be found in Ref.\,\cite{Jensen2017}.

The spin Hamiltonian for the electron-electron interaction $H_{\rm{ee}}$ can be written as: 
\begin{equation}
\label{eq:spin-hamiltonian}
 H_{\rm{ee}}=(D+d^{||}_{gs})S_z^2+d^{x}_{gs}(S_x^2-S_y^2)+d_{gs}^y(S_xS_y+S_yS_x)\,,
\end{equation}
where $d^{||}_{gs}$ and $d^x_{gs},d^y_{gs}$ are the electric field NV$^-$ coupling constants. The temperature dependence of $D$ is described by a fitting formula\,\cite{CambriaPRB2023} assuming two relevant phonon modes,
\begin{equation}
\label{eq:T dependence of D}
\begin{split}
D(T)=D_0+c_1n_1+c_2n_2\,,
\end{split}
\end{equation}
where $D_0$ is the ZFS at zero temperature, and $c_i$ describes how the corresponding phonon mode shifts the ZFS by modulating the
mean positions and vibrational amplitudes of the atoms
in the lattice, $c_1=-54.91$\,MHz and $c_2=-249.6$\,MHz respectively. $n_i$ is the mean occupation number of the $i$th mode,
\begin{equation}
\label{eq:mean occupation number}
\begin{split}
n_i=\frac{1}{e^{\Delta_i/k_BT}-1}\,,
\end{split}
\end{equation}
where $\Delta_i$ is the mode energy with $\Delta_1=58.73$\,meV and $\Delta_2=145.5$\,meV, respectively and $k_B$ is the Boltzmann constant. Assuming that $D$ is the only temperature-dependent parameter in the Hamiltonian and replacing $D$ by $D(T)$ in the spin Hamiltonian, we can get the frequency shifts of all the cross-relaxation features at different temperatures. The detailed discussion can be found in the Appendix.

\section{Experimental details} 
\subsection{Diamond samples}

We studied two diamond samples. The [111]-cut high-pressure high-temperature (HPHT) synthesised diamond sample S2 from Sumitomo with high density of color centers ([N]$\sim$100\,ppm, \cite{AcostaPRB2009}) was chosen because the features were best visible in it.
The second sample, [111]-cut CVD (Chemical Vapour Deposition) sample from Element-6 ([NV]$^-\sim$3.8\,ppm, $^{13}$C-depleted), was investigated only for the three main features (GSLAC, 591\,G and P1 center) to compare with the results obtained for the S2 sample.


\subsection{Experimental setup}

\begin{figure}
  \centering
\includegraphics[width=\columnwidth]{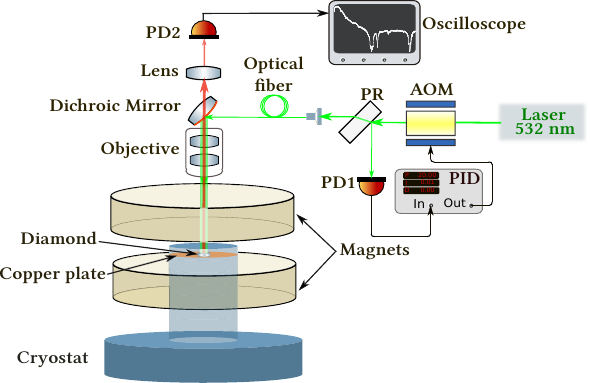}
\caption {Schematic of the experimental setup (PR: partial reflector; AOM: acousto-optic modulator; PD: photodiode; PID: proportional–integral–derivative controller). Note that some parts of the experimental apparatus are not shown in the figure, such as the drivers for the laser and AOM, transmission pipes for the cryostat, dewar with liquid helium and the cooling system for the magnets.}
\label{fig:setup} 
\end{figure}

The schematic of the experiment apparatus is shown in Fig.\,\ref{fig:setup}. The diamond sample is placed on a copper plate that is subsequently arranged on the top of a cold finger in a helium flow-through cryostat (Janis ST-500) mounted on the surface of an optical table. The copper plate has good thermal conductivity in a wide range of temperatures (4\,K-300\,K), in addition, it is cut to reduce eddy currents. The magnetic field is generated with two electromagnets in an arrangement designed to maximize the field homogeneity and minimize thermal drifts. The two magnets are mounted on a rotation stage (not shown in Fig.\,\ref{fig:setup}), enabling equal rotation of both magnets to change the projection of the magnetic field on the $z$ axis. The vertical position of the magnets as well as the spatial separation between them is fixed so that the diamond is located precisely in the middle, a region with optimal magnetic homogeneity. The accessible range of the magnetic field generated by the setup is between 0 to over 1024\,G (GSLAC feature region). 

Light with a wavelength of 532\,nm emitted by a CW laser (Laser Quantum GEM 532) goes through an acousto-optic modulator (AOM) and is guided with a partial reflector (PR). The reflected beam is used to monitor the output power of the laser and stabilize the intensity using a proportional–integral–derivative (PID) controller. The transmitted light is coupled into an optical fiber. The diverging beam from the output of the fiber passes through an objective and is focused on the diamond. The objective is mounted on a motorized 3D stage (not shown in Fig.\,\ref{fig:setup}) so we can optimize the parking spot of the beam on the diamond and obtain the maximum photoluminescence. The photoluminescence is then directed through a dichroic mirror and focused on the photodiode with a lens. 
 
The temperature around the diamond can be controlled within an uncertainty of $\pm\,0.1$\,K in a range of 300\,K-4\,K, using a temperature controller (LakeShore 335). In practice, when the temperature stabilized at one setting point, we collected the magnetic spectrum of the photoluminescence by sweeping the current supplied to the magnets. The magnetic shifts of the cross-relaxation features were calculated by comparing the position of features in the spectra acquired at different temperatures. The results are shown in Fig.\,\ref{fig: temperature dependence}.  

\section{Results}

\subsection{Temperature dependence of the cross-relaxation features} 

\begin{figure}
  \centering
\includegraphics[width=\columnwidth]{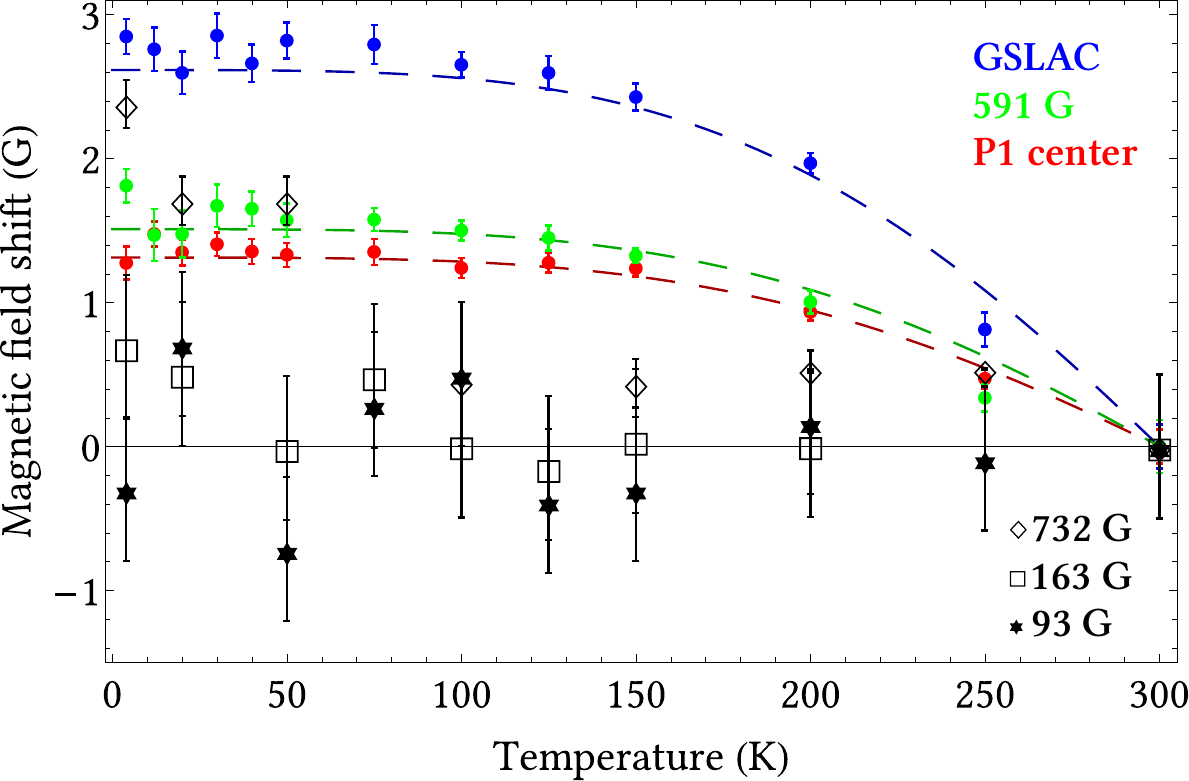}
\caption 
{ \label{fig: temperature dependence} 
Temperature dependence for all investigated features (S2 sample). Dashed lines represent theoretical predictions for the features' shift in the magnetic field (blue for GSLAC, green for 591\,G and red for P1 center), while full points represent experimentally obtained results. Additional features, the so-called small features, are represented by black markers. The shift of those in the magnetic field is very interesting, because with a decrease in temperature for two of them (93\,G and 163\,G) it is equal to 0. This phenomenon opens a wide range of possible applications of 93\,G and 163\,G features. Moreover, the 732\,G feature was discovered and even though we do not know its origin and therefore cannot calculate theoretical predictions for the shift in the magnetic field, it is clear that it is not zero like it was for the previous features.}
\end{figure}

We designate the magnetic spectrum acquired at room temperature as a baseline, and then all the magnetic spectra at different temperatures are compared with it. The magnetic shifts are determined for the six cross-relaxation features, which are presented in Fig.\,\ref{fig: temperature dependence}. Of the prominent features at 512\,G, 591\,G and 1024\,G, respectively, the experimental results agree well with the theoretical predictions for both diamond samples. In Fig.\,\ref{fig: temperature dependence}, only the results with the S2 sample are shown as colored dots, and the dashed lines are calculated predictions. 

Of particular interest are the so-called  ``small features'' at 163\,G, and 93\,G, respectively. These features appear to have small shift as a function of temperature, the black symbols in Fig.\,\ref{fig: temperature dependence}, although the uncertainties are larger compared to those for more prominent features. One may notice the increasing data point fluctuation in the low-temperature region (below 150\,K). This is because of the drop in photoluminescence at low temperatures making it harder to determine the center position of the profile. Despite that, the experimental data agree well for all features with theoretical predictions. 

The insensitivity of particular cross-relaxation features to temperature highlights promising potential applications. It enables the decoupling of the sensitivity of NV centers to temperature and to other environmental parameters, such as electric field, magnetic field, stress, etc., and the development of sensors immune to temperature fluctuation of the environment. 

Furthermore, the magnetic spectrum of the S2 sample presents a feature at 
732\,G
 as shown in Fig.\,\ref{fig:plT_2samples}, which we assign to a level crossing of a coupled system of two off-axis NV$^-$ centers and a P1 center, see Table\,\ref{Tab:predcition} and Appendix~B. 
The temperature shift for this feature does not appear to follow the general behavior seen for other features; however, more experiments would need to be done to make quantitative conclusions about this relatively small feature. Relying on our tentative assignment of the feature, we suspect that the structure of the 732\,G feature is not always resolved in our measurements, making it hard to identify the temperature shift of the feature precisely.

\subsection{Hyperfine structure}
\begin{figure}
  \centering
\includegraphics[width=\columnwidth]{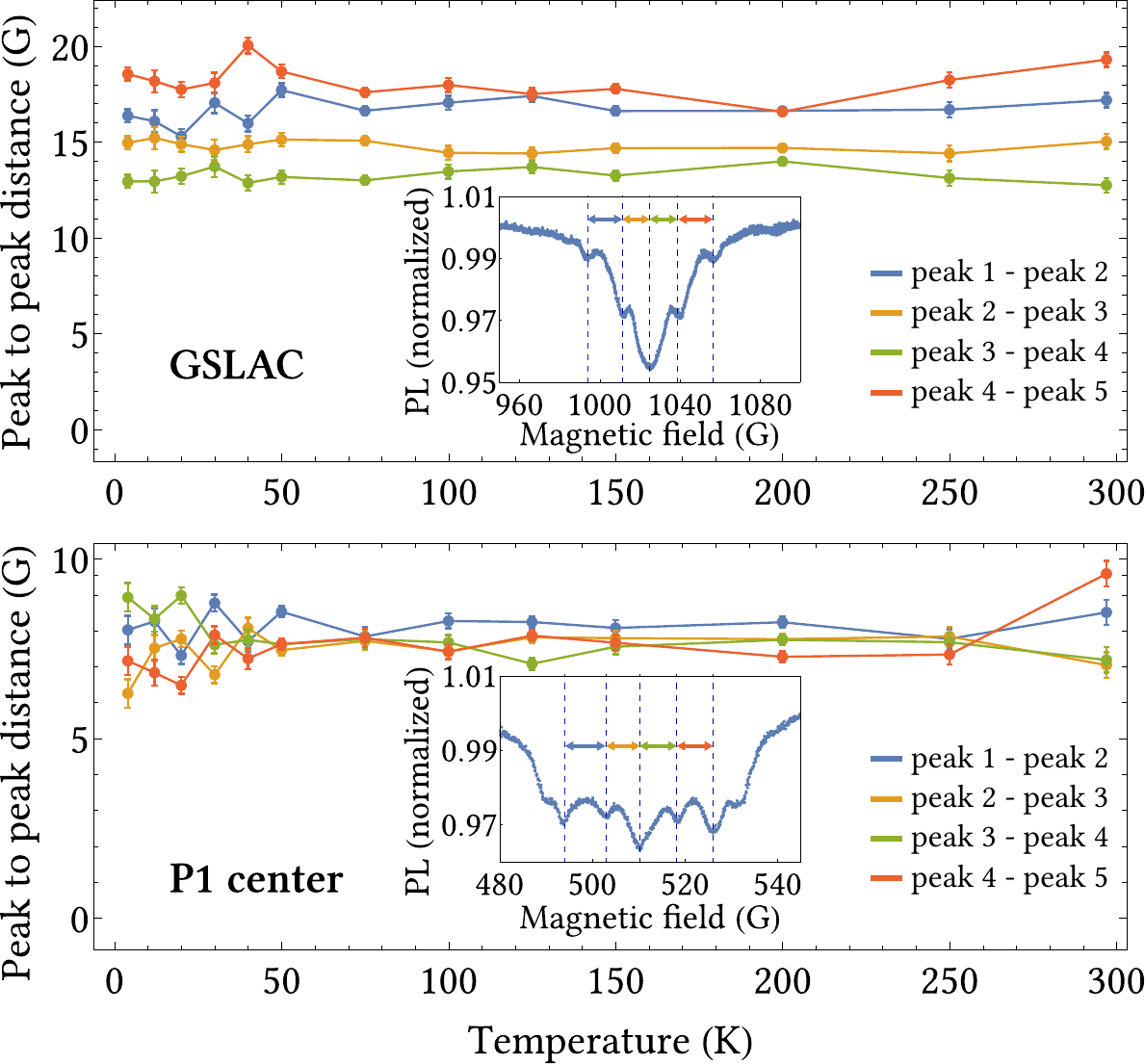}
\caption 
{Temperature dependence of the distance between the hyperfine structure side peaks at GSLAC \& P1 center features (S2 sample). Inserted plots represent the detailed structure of the features. 
}
\label{fig:hyperfine}
\end{figure}

As part of the analysis, we measured the separations and relative amplitudes between the substructure peaks in the cross-relaxation features and found that the distances between individual peaks and their relative amplitudes for a given feature did not significantly change with temperature. This is shown for the separations in Fig.\,\ref{fig:hyperfine}. This is important for data processing because when we cannot resolve the substructure due to, for example, poor signal-to-noise ratio, we can assume that the lineshape does not change with temperature to produce a systematic effect.

\section{Discussion \& Conclusion}

We report on the identification and temperature dependence of cross-relaxation features in the magnetic spectrum for NV centers in diamonds. 
Theoretically, we investigate the magnetic field dependence of the energy levels and look for (avoided) level crossings of various composite systems of NV$^-$ centers, P1 centers, and $^{13}$C nuclear spins.  We identify the causes for all the features except the low-filed features at 93\,G and 163\,G.
 By taking into account the temperature dependence of ZFS, the magnetic shifts of all features can be theoretically predicted. We carried out two measurements with two different diamond samples, E6 and S2. For the prominent features at 512\,G, 591\,G, and 1024\,G, the experimental results agree well with the theoretical predictions for both samples.

In addition, we discovered that two small cross-relaxation features, at 93\,G and 163\,G, seen with the S2 sample do not show a significant shift with temperature. The immunity of these features to temperature fluctuations in the environment may be of significance to NV-based quantum sensing technology. We also present magnetic shifts of the 
feature at 732\,G, which exhibits distinctive behavior compared with other features.

\section{Acknowledgements}
We thank Marcus Doherty, Andrey Jarmola, Till Lenz, Sean Lourette, Patrick Maletinsky, Matthew Markham, Mariusz Mr\'ozek, and Dieter Suter for informative discussions and helpful advice. This work was supported by the EU, project HEU-RIA-MUQUABIS-101070546, by the DFG, project FKZ: SFB 1552/1 465145163 and by the German Federal Ministry of Education and Research (BMBF) within the Quantumtechnologien program via the DIAQNOS project (project no. 13N16455). X. Z. was supported by Beijing 
National Laboratory for Condensed Matter Physics (2023BNLCMPKF013), National Natural Science Foundation of China (project no. 12404550), and by the Fundamental Research Funds for the Central Universities (DUT24RC(3)016). H.Z. acknowledges the support of Beijing Natural Science Foundation (L233021).
V.I. was supported by the National Research, Development and Innovation Office of Hungary (NKFIH) within the Quantum Information National Laboratory of Hungary (Grant No. 2022-2.1.1-NL-2022-00004) and project FK 145395. 
V.I. also acknowledges support from the Knut and Alice Wallenberg Foundation through the WBSQD2 project (Grant No.\ 2018.0071).


\section*{Appendix A}

The Zeeman Hamiltonian $H_z$ can be written as
\begin{equation}
\label{eq:Zeeman Hamiltonian}
\begin{split}
H_z=\mu_B g_s \boldsymbol{B} \cdot \boldsymbol{S}\,,
\end{split}
\end{equation}
where $\mu_B\approx1.4$\,MHz/G is the Bohr magneton, $g_s\approx 2$ is the Landé g-factor for the electron, and $\boldsymbol{B},\,\boldsymbol{S}$ represent the magnetic field and electron-spin operator, respectively. 

The Hamiltonian of the interaction between the electron spin $S$ and nuclear spin $I$ (which can be that of either $^{14}$N or $^{15}$N) can be expressed as 
\begin{equation}
\label{eq:nuclear Hamiltonian}
\begin{split}
H_N= \boldsymbol{S} \cdot \mathbf{A}_N \cdot \boldsymbol{I} + \gamma_N \boldsymbol{B} \cdot \boldsymbol {I} +  \boldsymbol{I} \cdot\,\mathbf{Q}\,\cdot\,\boldsymbol{I}\,,
\end{split}
\end{equation}

where $\mathbf{A}_N$ represents the hyperfine interaction tensor, $\gamma_N$ is the nuclear gyromagnetic ratio, and $\mathbf{Q}$ denotes the nuclear-quadrupole-splitting tensor.

The Hamiltonian of the P1 center, which has electronic spin $S_{\rm{P1}}=1/2$ and nuclear spin $I_{\rm{P1}}=1$ (in the case of $^{14}$N), is 

\begin{equation}
\label{eq:P1 Hamiltonian}
\begin{split}
H_{P1} =& \boldsymbol{S}_{\rm{P1}} \cdot \mathbf{A} \cdot \boldsymbol{I}_{\rm{P1}} 
+ \gamma_e \boldsymbol{B} \cdot \boldsymbol{S}_{\rm{P1}} 
- \gamma_I \boldsymbol{B} \cdot \boldsymbol{I}_{\rm{P1}} \\&+ \boldsymbol{I}_{\rm{P1}} \cdot\,\mathbf{Q}_{\rm{P1}}\,\cdot\,\boldsymbol{I}_{\rm{P1}}\,
\end{split}
\end{equation}

where $\mathbf{A}$ is the hyperfine tensor which is diagonal when the quantization axis is parallel to the symmetry axis of the P1 center, $\mathbf{A}$=$\begin{pmatrix}
A_{\perp} & 0    & 0\\
0    & A_{\perp} & 0\\
0    & 0    & A_{\parallel}
\end{pmatrix}$
with $A_{\perp}=81.3$\,MHz and $A_{\parallel}=114$\,MHz\,\cite{Kongnatcomm2018}, 
$\gamma_e$ is the electron gyromagnetic ratio, $\gamma_I$ is the nuclear gyromagnetic ratio, and $\mathbf{Q}_{\rm{P1}}$ denotes the nuclear-quadrupole-splitting tensor for P1 centers \cite{Jensen2017}.

To predict the positions of cross-relaxation features, we calculate the eigenvalues of the respective Hamiltonians as a function of the magnetic field and search for the cross points in the transition energies, as shown in the figures in Table\,\ref{Tab:predcition}. The frequency (magnetic) shifts of those features when temperature changes are calculated by taking the difference in the eigenvalues by plugging $D(T)$ in the spin Hamiltonian.

\section*{Appendix B}

Here we briefly describe the methodology used to identify allowed and avoided crossings of the energy levels of finite dimensional multi-spin systems. The literature uses different terminologies and pictures to describe relaxation phenomena of multi-spin systems. Line crossings and cross-relaxation features of distinct spins can also be discussed in terms of energy level crossing and avoided crossing in the composite Hilbert space of the spins. The fact that two spins have a resonance in their transition energies equally means that the energy-level structure of the composite system exhibits a degeneracy. When cross-relaxation occurs between the spins, the degeneracy is lifted by a coupling Hamiltonian term that mixes the states of the two spins. Hereinafter, we study composite systems and look for relevant energy level (avoided) crossings to identify magnetic field values of actual and potential features in the magnetic field-dependent photoluminescence (PL) signal. 
To this end, we construct the Hamiltonian $H$ of a coupled spin system and diagonalize it. By shifting the diagonal element of the Hamiltonian, we make sure to have all eigenvalues in the positive domain at all considered values of the magnetic field $B$. 
This enables us to order the eigenstates $\psi_i$ using the absolute values of the eigenenergies $\varepsilon_i$ and avoid unwanted changes in the ordering when scanning the magnetic field that can give rise to false signals in our methodology.
To look for (avoided) crossings that include the $\ket{m_S=0}$ state of the NV center, we compute the projection $p_i (B) = | \braket{m_S=0}{\psi_i (B)} |^{2}$ and plot against the value of the magnetic field applied along the quantization axis. Here, we are particularly interested in crossings that involve the $\ket{m_S=0}$ state since the mixing of these states with $\ket{m_S=\pm 1}$ can lead to features in the photoluminescence signals. Crossings of the $\ket{m_S= +1}$ and $\ket{m_S= -1}$ states are omitted in  our analysis. 

\begin{figure}
  \centering
\includegraphics[width=0.85\columnwidth]{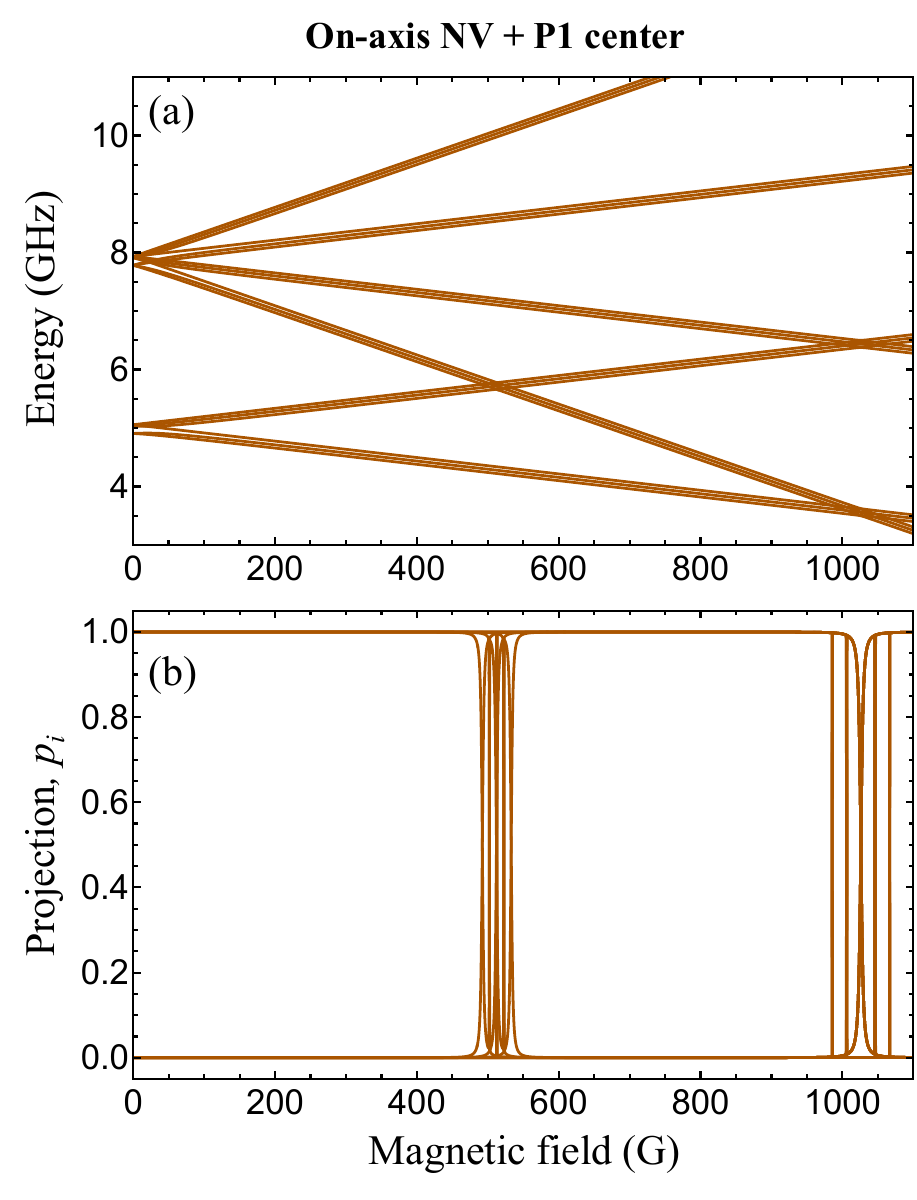}
\caption 
{ \label{fig:nv-p1} 
Energy-level structure (a) and state-vector projections (b) for a three-spin NV-P1-$^{15}$N system as a function of the applied magnetic field. }
\end{figure}

As an example, we discuss the case of a coupled on-axis NV center and P1 center with a $^{14}$N hyperfine interaction included. Figure \ref{fig:nv-p1} a) depicts the magnetic field dependence of energy eigenstates $\varepsilon_i$. The energy levels exhibit crossings and avoided crossings at different values of the external magnetic field. In Fig.\,\ref{fig:nv-p1} b) one can see the computed values of the $p_i$ projections as a function of the magnetic field. The $p_i$ values drop from 1 to 0 at certain magnetic field values signaling relevant crossing and avoided crossings of the states $\ket{m_S=0}$ and $\ket{m_S=-1}$. Note that the crossings of the $\ket{m_S=\pm 1}$ manifold are not indicated by the drop of the $p_i$ values. For instance, close to zero magnetic fields, the energy levels do exhibit crossing in the $\ket{m_S=\pm 1}$ manifold, however, since these crossings do not influence the $\ket{m_S = 0}$ contribution of the eigenstates, no change can be observed in the $p_i$ values. At the GSLAC and around halfway between the GSLAC and $B=0$, we observe significant variation of the $p_i$ values indicating the presence of relevant crossing and avoided crossings. 

At the GSLAC we can see two types of features in the magnetic field dependence of the $p_i$ values. There are four straight vertical lines close to the GSLAC as well as a Lorentzien-like peak pluss dip structure touching each other at the magnetic field value of the GSLAC. The former indicates true crossings while the latter indicates avoided crossings. The true crossings observed in Fig.\,\ref{fig:nv-p1} may also be relevant for cross-relaxation processes, since additional perturbations, such as off-axis magnetic field, strain, and nuclear spins may enable mixing of the states beyond what is described by our Hamiltonian and lead to features in the PL signal. Indeed, GSLAC features as well as the four satellites have been experimentally observed and studied theoretically \cite{Auzinsh2019GSLAC,Ivady2021PRB} and observed in our measurements, see Fig.\,\ref{fig:plT_2samples} and Fig.\,\ref{fig:hyperfine}. The characteristic P1 cross relaxation feature at around 512\,G is also indicated by the change of the $p_i$ values.

\begin{figure}
  \centering
\includegraphics[width=0.85\columnwidth]{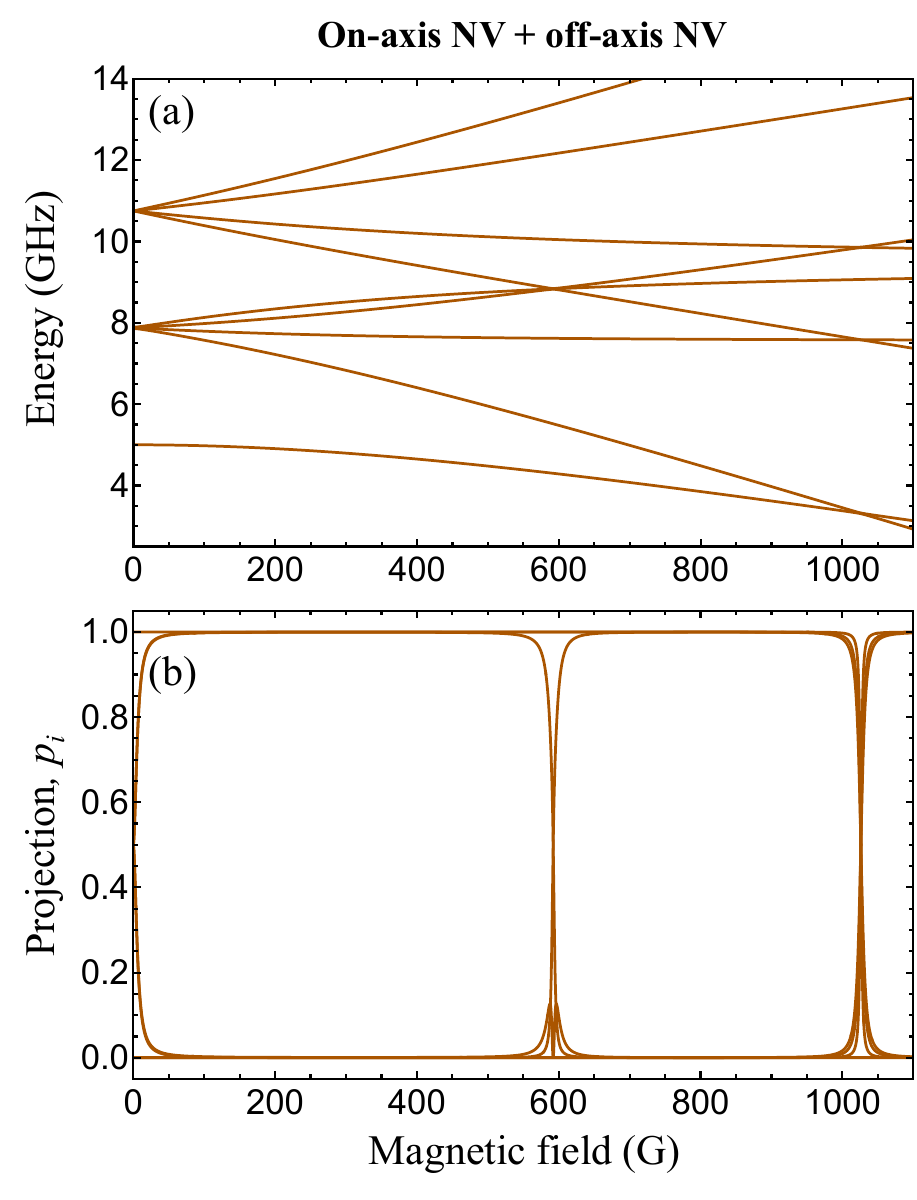}
\caption 
{ \label{fig:nv-onv} 
Energy-level structure (a) and state-vector projections (b) for an on-axis NV center - off-axis NV center system as a function of the applied magnetic field.}
\end{figure}

As another example, we depict the energy levels and the $p_i$ values for a pair of non-collinear NVs as a function of the external magnetic field parallel to the symmetry axis of one of the NV centers in Fig.\,\ref{fig:nv-onv}. We can observe three avoided crossings and spin mixing features, one at the GSLAC, one at 591\,G, and one at 0 magnetic field. The latter indicates the so-called zero-field feature observed in NV ensembles, see \cite{Dhungel2024_Near_zero} and references therein.

Finally, we note that our method is not indicative of the depth of the dips caused by the different spin-relaxation features. For this, dynamic simulations can be run once the origins of the features are understood.   

\section*{Appendix C}

In this Appendix, we examine NV systems including different spins, such as on- and off-axis NV-centers, P1 centers, nitrogen, and carbon nuclear spins, up to five-spin clusters. The nuclear spins interact with their corresponding electron spin through the hyperfine coupling tensors (established in the literature). For simplicity, we omit the nitrogen spin of the NV center, however, include the nitrogen of the P1 centers, which can give rise to resolvable hyperfine or superhyperfine structures. For the electron spin-electron spin coupling, we use ad hoc, sometimes overestimated, coupling terms to identify the nature of the crossings. This approach does not alter our statements, however, it helps us to separate true crossings from avoided crossings. To avoid a shift of the crossing point, we keep the secular component of the coupling tensor low. In the following, we discuss relevant few-spin systems that either explain unknown PL features or support previous identifications.  

\begin{figure}
  \centering
\includegraphics[width=0.85\columnwidth]{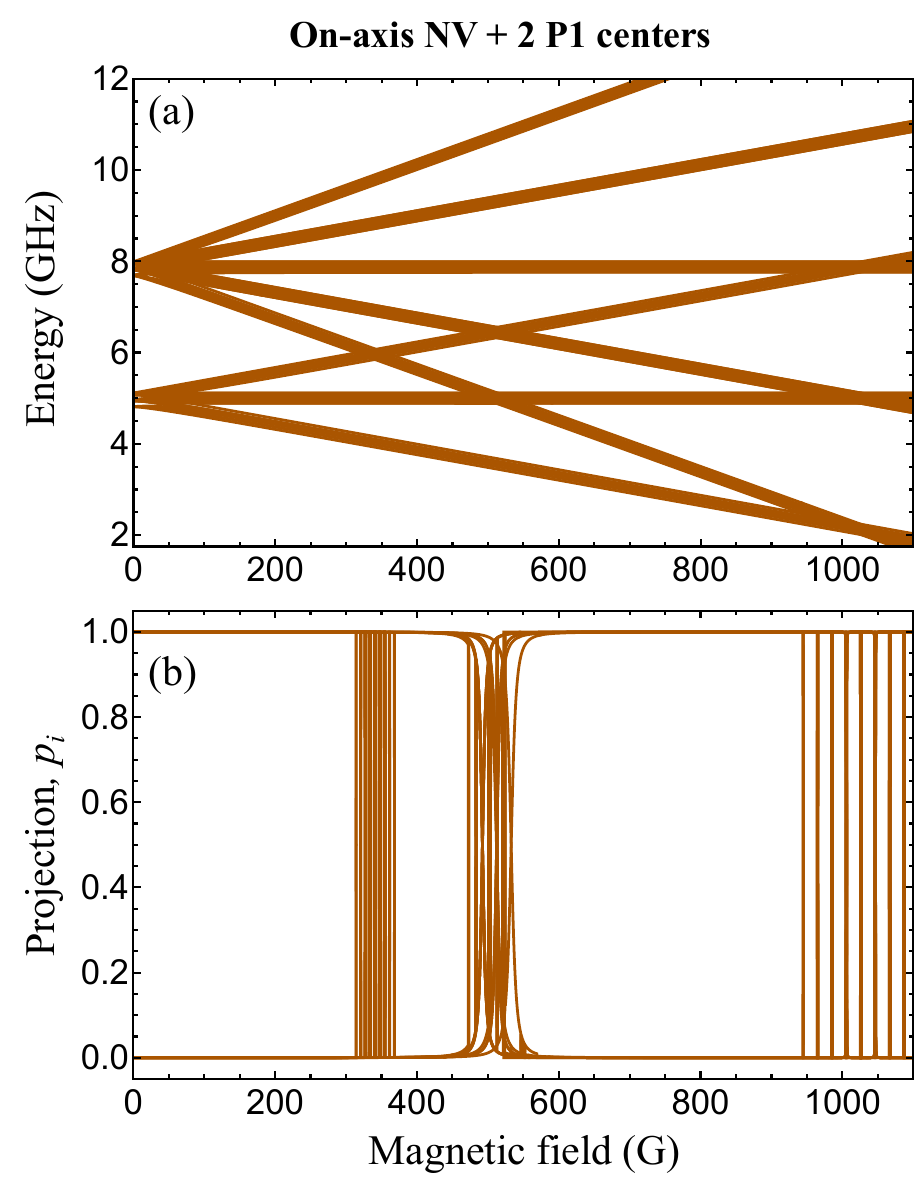}
\caption 
{ \label{fig:nv-2p1} 
Magnetic field dependence of the energy level structure and state vector projections for an on-axis NV center interacting with two P1 centers.}
\end{figure}

\emph{On-axis NV center interacting with two or three P1 centers.} As can be seen in Fig.\,\ref{fig:nv-2p1}, the presence of two interacting P1 centers can give rise to an additional broad resonance feature at around 342\,G. Considering a $^{14}$N hyperfine coupling tensor with a principal axis parallel to the axis of the NV center the crossing feature can extend from 314\,G up to 368\,G and include numerous lines.  These results are in remarkable agreement with the experimental observations reported recently in Ref.\,\cite{Wunderlich2021}. 

The inclusion of a third P1 center, not shown, can give rise to an additional broad feature centered around 257\,G. Both of these features can be seen in Fig.\,\ref{fig:plT_2samples}(b), although the detailed structure and the center of the peaks cannot be resolved and identified.

\begin{figure}
  \centering
\includegraphics[width=0.85\columnwidth]{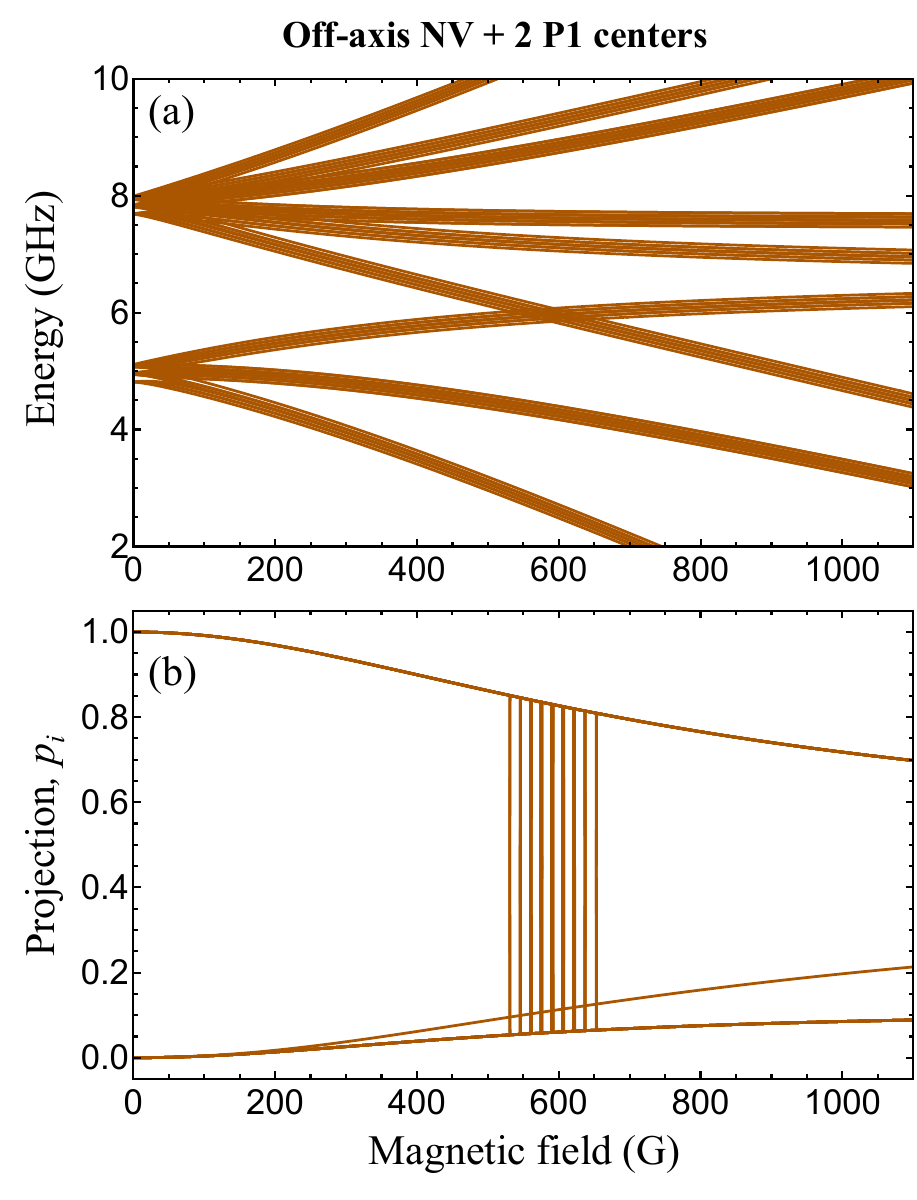}
\caption 
{ \label{fig:onv-2p1} 
Magnetic field dependence of the energy level structure and state vector projections for an off-axis NV center interacting with two P1 centers.}
\end{figure}

\emph{Off-axis NV center interacting with two or three P1 centers.} Figure\ref{fig:onv-2p1} depicts the energy eigenstates and the  $p_i$ values as a function of the magnetic field. Interestingly, numerous crossings including the $\ket{m_S =0} $ state are revealed. There are altogether nine separate lines observed, the central line located at 591\,G. This feature may contribute to the 591\,G feature often observed in experiments and assigned to the on-axis NV - off-axis NV system, however, due to the mixing of the NV states the PL signature of the off-axis NV - two P1 system may be suppressed. 

When a third P1 center is added to the system, not shown, two additional crossing regions appear at lower magnetic fields. The centers of the potential cross-relaxation features are at 347\,G and 497\,G.

\begin{figure}
  \centering
\includegraphics[width=0.85\columnwidth]{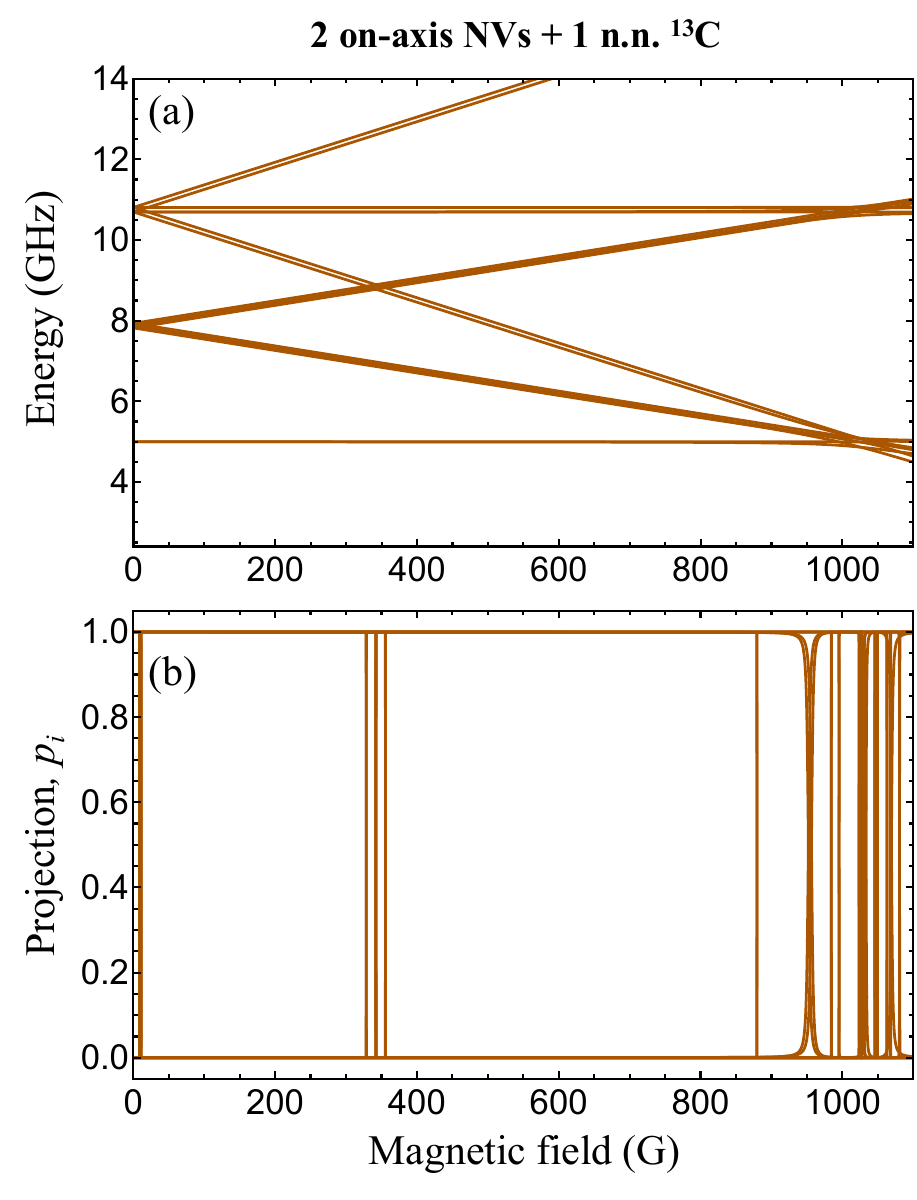}
\caption 
{ \label{fig:2nv-13C} 
Magnetic field dependence of the energy level structure and state vector projections for two on-axis NV centers interacting with a nearest-neighbor $^{13}$C nuclear spin.}
\end{figure}

\emph{Two interacting on-axis NV centers with and without one first neighbor $^{13}$C nuclear spin.} 
As can be seen in Fig.\,\ref{fig:2nv-13C}(a), the energy levels of two on-axis NV centers also exhibit a crossing at 342\,G. This effect is referred to as NV-NV auto cross-relaxation in Ref.\,\cite{Wunderlich2021} and is mentioned as the main source of the PL features observed at 342\,G. Due to the lack of always present strong nuclear spin interactions, however, auto cross-relaxation of aligned NV centers cannot account for the broad feature with several distinct dips centered around 342\,G. Presumably, both NV-2$\times$P1 systems and two-NV systems contribute simultaneously to the broad feature at 342\,G seen in our experiments as well as in Ref.\,\cite{Wunderlich2021}. 

\begin{figure}[H]
  \centering
\includegraphics[width=0.85\columnwidth]{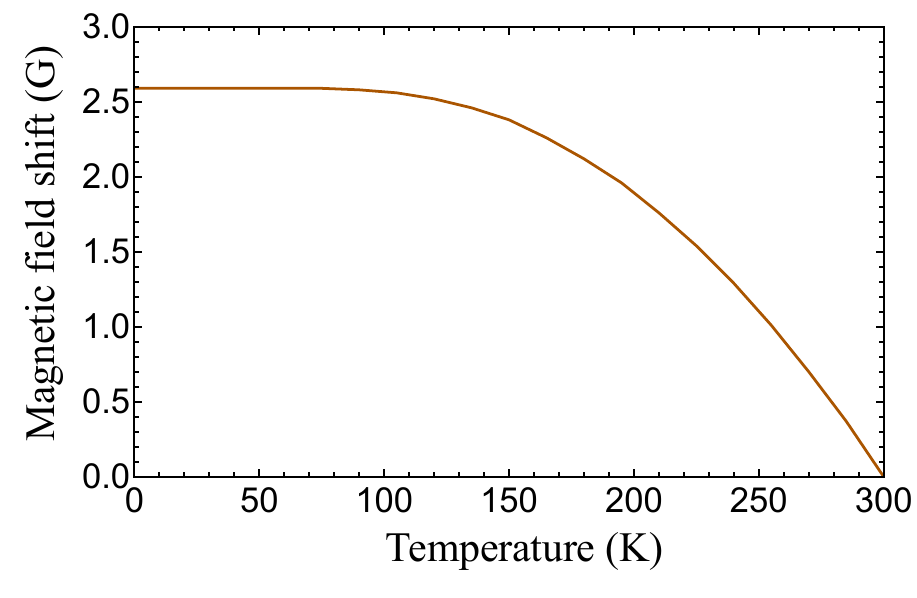}
\caption 
{ \label{fig:956-T-dep} 
Temperature-dependent magnetic field shift of the 956\,G crossing of the two interacting on-axis NV centers with a first neighbor $^{13}$C nuclear spin.} 
\end{figure}

 When a single $^{13}$C nuclear spin from the first shell of one of the NV center is included in the system, we observe split lines $\sim$13\,G away from the central line at 342\,G, however, this cannot account for the breadth of the feature at this magnetic field value. Interestingly, we observe a $\sim$1\,G shifted zero-field feature as well as new crossings at the left side of the GSLAC indicated by the  $p_i$ values. The notable ones are the true crossing at 879\,G and an avoided crossing at 952-956\,G. 
 It is important the note concerning these new crossings close to the GSLAC, that they are not implicitly GSLAC features. The true and avoided crossings at 879\,G and 952-956\,G happen within a single branch of the energy levels, which include both $\ket{m_S =0}$ and $\ket{m_S = \pm 1}$ states of the two NV centers, and not at the GSLAC crossing of two branches. Finally, the predicted temperature shift of the 956\,G feature is depicted in Fig.~\ref{fig:956-T-dep}.

\begin{figure}
  \centering
\includegraphics[width=0.85\columnwidth]{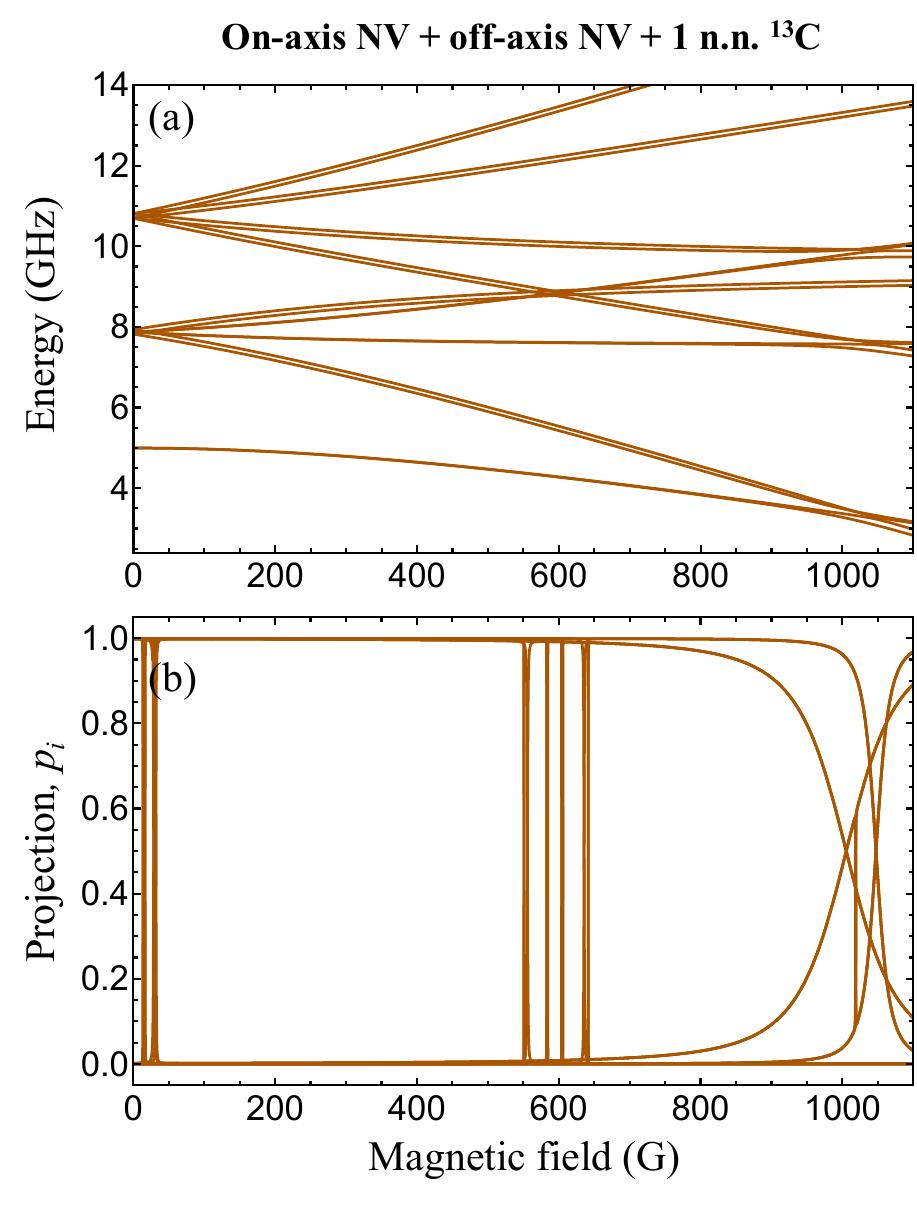}
\caption 
{ \label{fig:onv-13C} 
Magnetic field dependence of the energy level structure and state vector projections for interacting on-axis and off-axis  NV centers including a first neighbor $^{13}$C nuclear spin.}
\end{figure}

\emph{One an-axis and one off-axis NV center interact with one first neighbor $^{13}$C nuclear spin.} Previously, we considered the case of two on-axis NV centers and one $^{13}$C nuclear spin. By changing the symmetry axis of one of the NV centers, we again observe new crossings, see Fig.\,\ref{fig:onv-13C}. Close to the magnetic field of GSLAC, we find two avoided crossings at 1005~G and 1048~G. Furthermore, close to the 591~G feature, we find both a series of true crossing and avoided crossings at magnetic field values 551-556~G, 584~G, 606~G, and 636-642~G. The crossings may give rise to satellite  dips close to the 591~G feature. Finally, we find additional crossings at the zero field feature at around 4-7~G and 29-32~G some of which are allowed and could give rise to small dips at the zero field feature.

\begin{figure}
  \centering
\includegraphics[width=0.85\columnwidth]{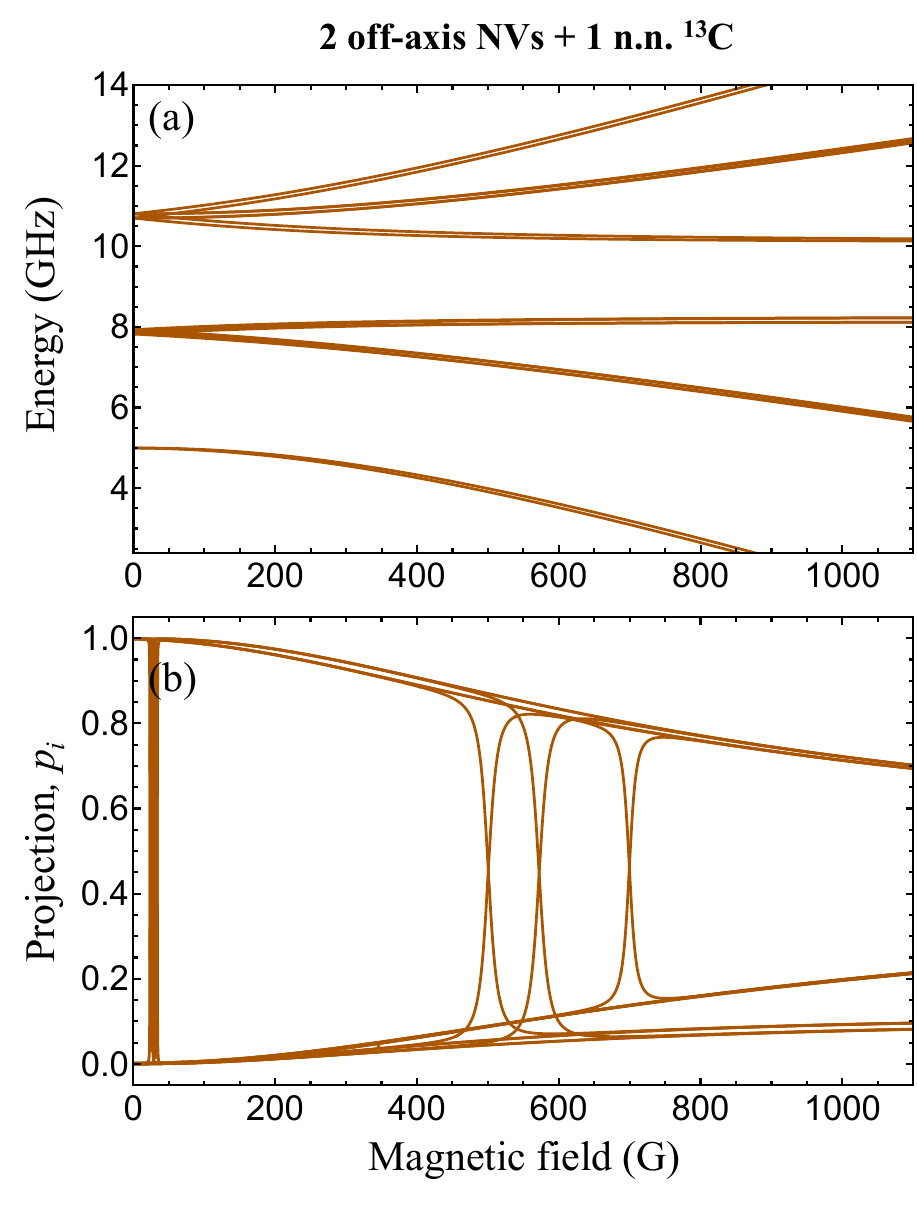}
\caption 
{ \label{fig:oonv-13} 
Magnetic field dependence of the energy level structure and state vector projections for two interacting off-axis  NV centers including a nearest-neighbor $^{13}$C nuclear spin.}
\end{figure}

\emph{Two off-axis NV centers interact with one nearest-neighbor $^{13}$C nuclear spin.}
Considering two off-axis NV centers, one of them includes a $^{13}$C nuclear spin, we observe three avoided crossings in the range of 500-700\,G, see Fig.\,\ref{fig:oonv-13}(b). In particular, we find the center of the avoided crossings at 501\,G, 572\,G, and 700\,G. The energy levels, however, do not show apparent crossings at these magnetic field values, see Fig.\,\ref{fig:oonv-13}(a), since these features are also related to intra-branch crossings. The former two overlap with other cross-relaxation features, while the latter could not be resolved in our experiments. In addition to these features, our method reveals numerous crossings in the 24-34\,G interval close to the zero-field feature.   

\begin{figure}
  \centering
\includegraphics[width=0.85\columnwidth]{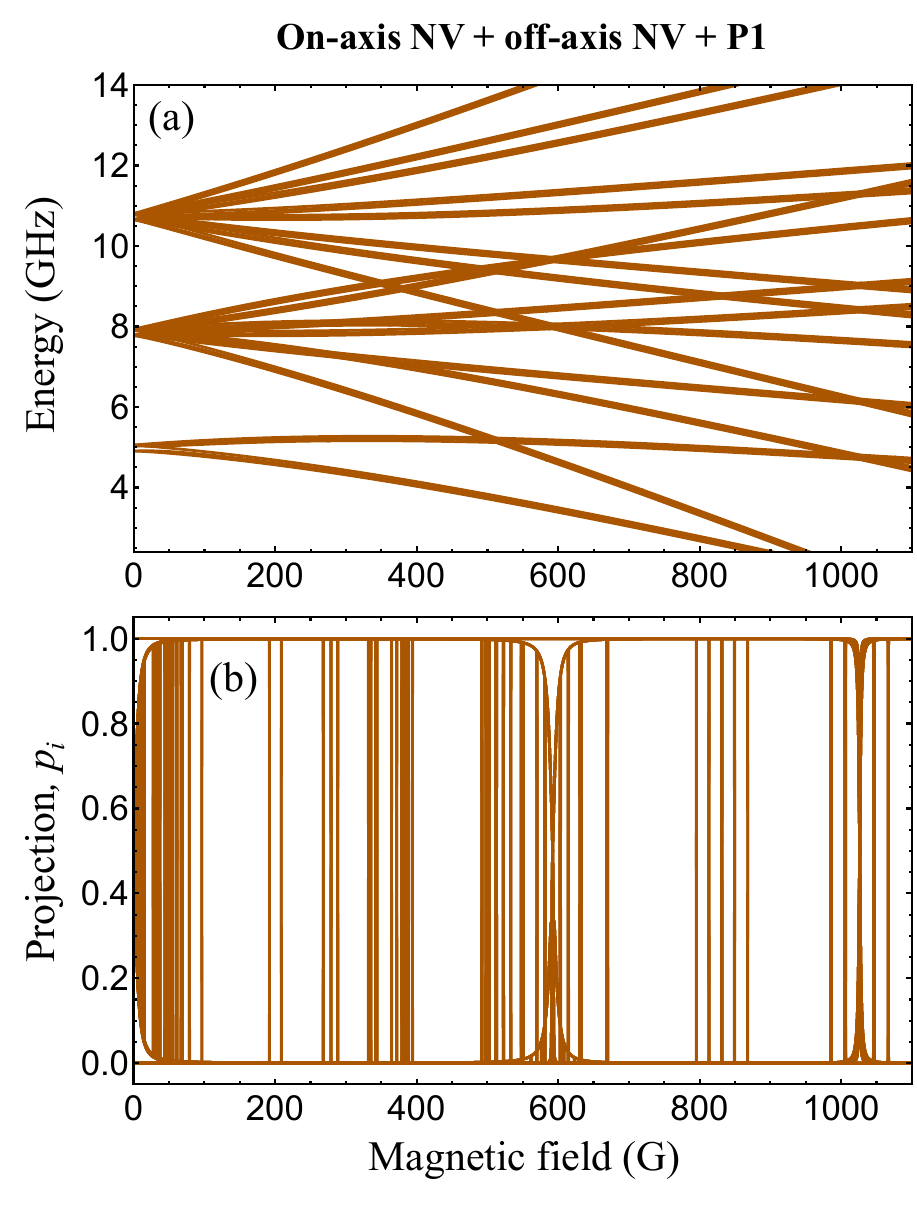}
\caption 
{ \label{fig:nv-onv-p1} 
Magnetic field dependence of the energy level structure and state vector projections for one on-axis and one off-axis NV centers interacting with a P1 center.}
\end{figure}

\emph{One on-axis and one off-axis NV center interact with one P1 center.} The most complicated energy level structure is observed for this system including three different electron spin and a $^{14}$N nuclear spin. In addition to the numerous apparent crossings, the  $p_i$ values indicate numerous other crossings within the bands. The crossings of different branches occur at 332-342~G, 365-394~G, 492-502~G, 795-868~G, and  at the GSLAC. The crossings close to the zero field feature spread from 0~G to 96~G. To judge the relevance of these crossings and to estimate the magnitude of related PL features spin dynamics simulations are needed, which is out of the scope of the present article.

\begin{figure}[H]
  \centering
\includegraphics[width=0.85\columnwidth]{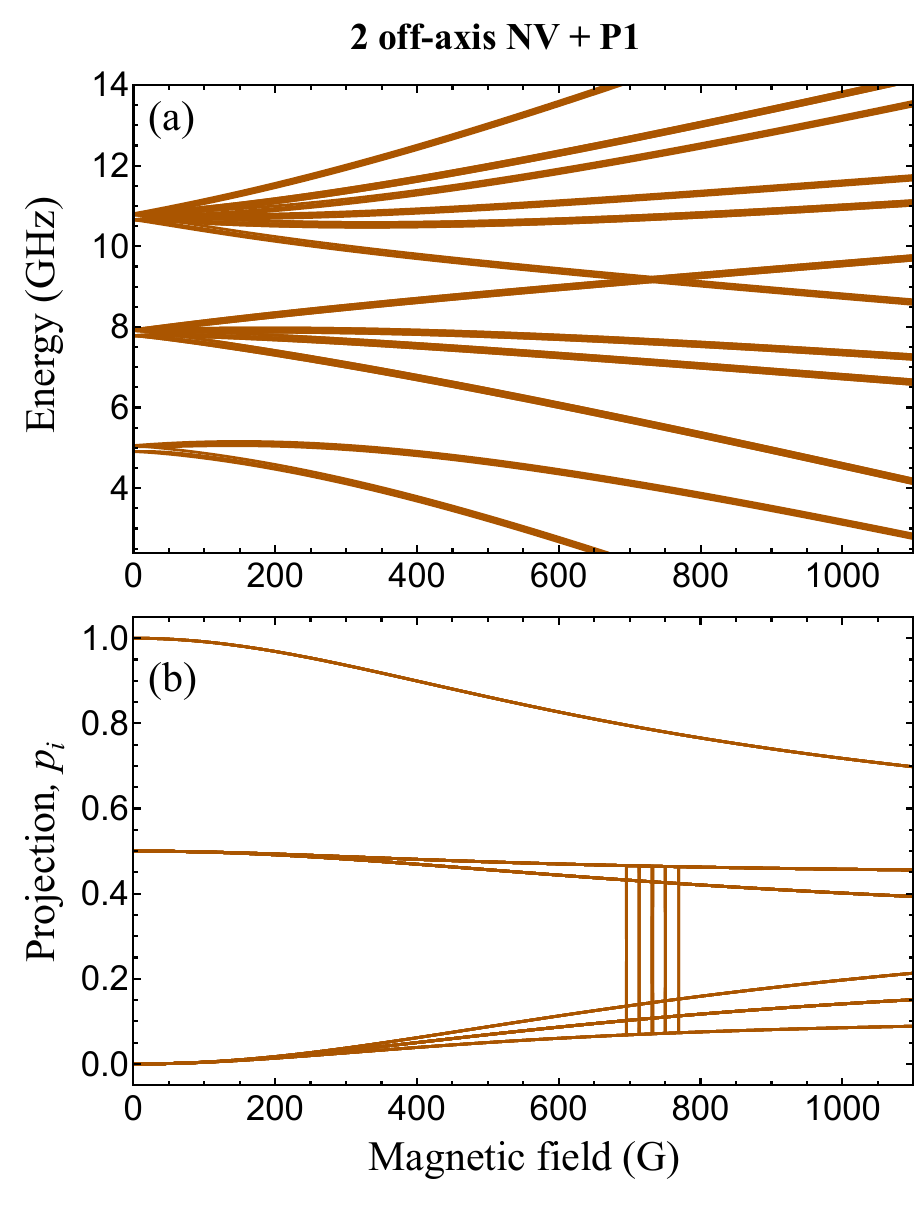}
\caption 
{ \label{fig:2onv-p1} 
Magnetic field dependence of the energy level structure and state vector projections for two off-axis NV centers interacting with a P1 center.}
\end{figure}

\begin{figure}[H]
  \centering
\includegraphics[width=0.85\columnwidth]{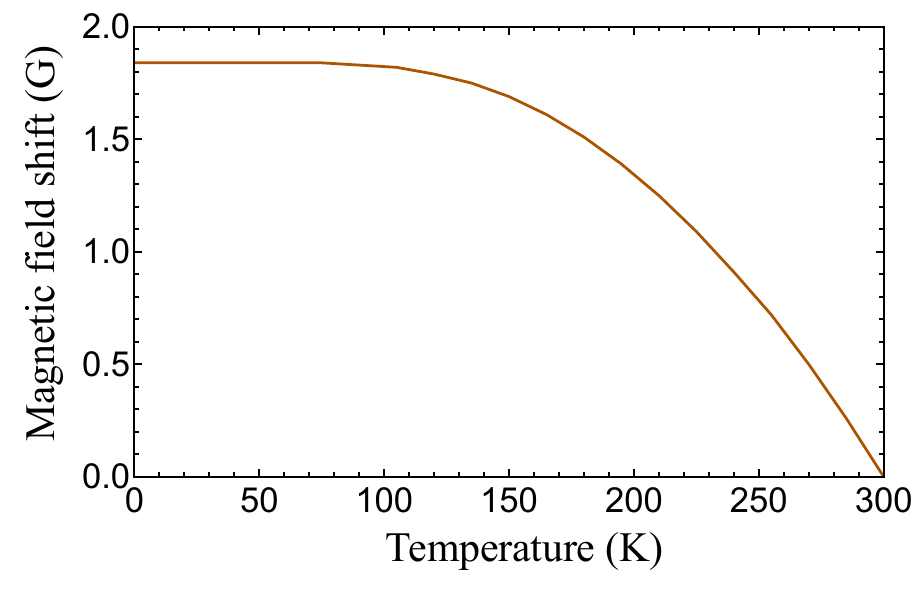}
\caption 
{ \label{fig:732-T-dep} 
Temperature-dependent magnetic field shift of the 732\,G crossing of the two off-axis NV centers-P1 center system.} 
\end{figure}

\emph{Two off-axis NV centers interact with a P1 center.}
The solution of the two off-axis NV centers and one P1 center is considerably simpler compared to the previous case and gives rise to only a single crossing feature centered around 732\,G with additional split crossing lines at 695\,G, 714\,G, 750\,G, and 769\,G. Interestingly, the central line perfectly matches the feature at 732\,G consistently observed in our temperature-dependent measurement. We tentatively assign this crossing to the observed PL feature and numerically studied the shift of the crossing point, i.e.,  we calculated the temperature shift of the 732\,G peak of the two off-axis NV centers, P1 center system assuming that it is solely caused by the change in $D$. As can be seen in Fig.\,\ref{fig:732-T-dep}, we observe a total of -1.84\,G shift of the feature when increasing the temperature from 0 to 300\,K. The magnitude of the shift is comparable with our measurements for the 732\,G feature, see Fig.\,\ref{fig: temperature dependence}. However, the experimental data  show a sudden change in the shift not reproduced in our simulations. A possible reason could be the predicted  structure of the 732\,G feature, which is not resolved in our measurement and makes the identification of the position of the feature more challenging.  

\bibliographystyle{apsrev4-2-2}
\bibliography{literature}

\end{document}